\documentstyle[12pt,aasms4]{article}

\begin{document}
\centerline{Submitted to the Astrophysical Journal}
\vskip 0.3in
\title{A Fit To The {\it Simultaneous} Broad Band
Spectrum Of Cygnus X-1 Using The Transition Disk Model}

\author{\bf R. Misra}
\affil{Inter-University Centre for Astronomy and Astrophysics, Pune, India}
\authoremail{rmisra@iucaa.ernet.in}
\author{\bf V. R. Chitnis}
\affil{Tata Institute of Fundamental Research, Homi Bhabha Road, Mumbai, India.}
\authoremail{vchitnis@tifrc3.tifr.res.in}
\author{\bf Fulvio Melia\altaffilmark{1}}
\affil{Department of Physics and Steward Observatory, University of Arizona,Tucson AZ}
\authoremail{melia@physics.arizona.edu}
\altaffiltext{1}{Presidential Young Investigator.}

\begin{abstract}
We have used the transition disk model to fit the {\it simultaneous}
broad band ($2-500$ keV) spectrum of Cygnus X-1 from  OSSE and
Ginga observations. In this model, the spectrum is produced by saturated
Comptonization within the inner region of the accretion disk, where 
the temperature varies rapidly with radius.  In an earlier attempt,
we demonstrated the viability of this model by fitting the data
from EXOSAT, XMPC balloon and OSSE observations, though these were
not made simultaneously.  Since the source is known to be variable,
however, the results of this fit were not conclusive.  In addition,
since only once set of observations was used, the good agreement
with the data could have been a chance occurrence.  Here, we improve
considerably upon our earlier analysis by considering {\it four}
sets of {\it simultaneous} observations of Cygnus X-1, using an 
empirical model to obtain the disk temperature profile.
The vertical structure is then obtained using this 
profile and we show that the analysis is self-consistent.
We demonstrate conclusively that the transition disk spectrum is a better fit 
to the observations than that predicted by the soft photon Comptonization
model. In particular, although the transition disk model has only one 
additional parameter, the $\chi^2$ value is reduced and there are no systematic
residuals.  Since the temperature profile 
is obtained by fitting the data, the 
unknown viscosity mechanism need not be specified. The disk structure 
can then be used to infer the viscosity
parameter $\alpha$, which appears to vary with radius and luminosity.
This behavior can be understood if $\alpha$ depends intrinsically on
the local parameters such as density, height and temperature. However,
due to uncertainties in the radiative transfer, quantitative statements
regarding the variation of $\alpha$ cannot yet be made. 
\end{abstract}


\section{Introduction}

During the last two decades, the X-ray/$\gamma$-ray spectrum of 
Cygnus X-1 has been extensively studied by, e.g., HEAO1, EXOSAT, 
Ginga and the Compton GRO. This black hole system settles into
two distinct long term configurations, namely the so-called soft and hard
states.  In the hard state, the X-ray spectrum is dominated by $2-200$ 
keV emission, which has been interpreted by some to be the result of 
the inverse Compton scattering of cold external photons by a hot
($kT\approx 50$ keV) low density plasma. This plasma could
be the innermost region of the accretion disk (Shapiro, Lightman 
\& Eardley 1976 ), or it could be a corona overlying a relatively 
cold disk (Liang \& Price 1977; Haardt \& Maraschi 1993). 
An alternative model is one in which the emission is a sum of 
multi-temperature Wien peaks from the transition region of the disk 
(Maraschi \& Molendi 1990; Misra \& Melia 1996). All these
models predict a similar spectrum which is more or less a
power-law with an exponential cutoff.

However, it may now be possible to differentiate between these models by
combining the data from various high-energy instruments, such as
OSSE, Ginga and EXOSAT, and thereby studying the broad band 
nature of this source. The initial steps in this direction were 
taken by Chitnis, Rao \& Agrawal (1997) and Gierlinski et al. 
(1997). They concluded that a simple soft-photon Comptonization 
model does not adequately fit the data. Instead, significant
systematic residuals were obtained, and it appeared that an additional 
{\it ad hoc} spectral component is needed to account for the 
observations. Gierlinski et al. (1997) have also pointed out 
that the spectrum is photon starved which argues against an
overlying coronal model in which roughly half of the radiated
high-energy photons are reflected back into the emitting region
by the cold disk.

Subsequently, Misra et al. (1997) showed that the transition disk 
model (Misra \& Melia 1996) appears to fit the broad band data
without invoking an additional component. In this picture,
the structure of the accretion disk is obtained without assuming
{\it a priori} that the effective optical depth in the vertical 
direction is much greater than unity. This leads to an electron 
temperature that varies rapidly with radius (Maraschi \& Molendi 1990; 
Misra \& Melia 1996), and the spectrum differs in a significant way
from a simple power-law with an exponential cut off. This spectrum 
fits the data well and is not associated with any systematic
residuals. These results indicated that the transition disk model
may be a better fit to the observations than the spectrum expected
from a single temperature external photon Comptonization. 

However, the earlier analysis by Misra et al. (1997) used 
non-simultaneous observations by EXOSAT, XMPC balloon flights 
and OSSE. Since the source is known to be variable, the results 
of this analysis were not conclusive.  Moreover, since only one 
set of observations was used, it was not clear whether the agreement
with the transition model was a chance occurrence.  An additional
complication was the fact that a direct $\chi^2$ comparison between 
the transition disk and soft photon Comptonization models
could not be made since the latter was not fit to the same 
data. It was shown only that the transition model did not produce
any systematic residuals whereas the Comptonization model (fit
to a different data set) did give rise to these. 

In this paper, we use four sets of simultaneous observations
of Cygnus X-1 made with Ginga and OSSE. These data have already 
been analyzed with an isotropic Comptonization model by Gierlinski et al. 
(1997). Here, we reanalyze the data using the transition disk 
model and compare our results with those of the previous study.
Following Misra et al. (1997) we fit the data to an empirical model
instead of making a direct fit to the more complicated exact equations.
The best fit temperature profile is then used to determine the disk
structure. In the final step, the viscosity parameter $\alpha$ is 
inferred from the disk structure.

In the next section, we briefly describe the empirical model developed
by Misra et al. (1997).  In \S 3 the observations and the results of the
spectral fit to the transition model are compared with those obtained
by Gierlinski et al. (1997).  In \S 4 we calculate the disk structure 
from the best fit temperature profile and we discuss the self-consistency 
of our analysis. \S 5 summarizes the basic results.

\section{The Empirical Model}

To keep the calculation tractable, we use an empirical model
developed by Misra et al. (1997) to fit the data, rather than
attempt a solution of the full equations. We emphasis that the difference between the transition disk model and 
the empirical model is in the parametrization. Instead of the physical
parameters such as the mass accretion rate, the black hole mass and 
the viscosity parameter $\alpha$, we use empirical parameters like the 
temperature profile and the absolute normalization described below. 

The
gravitational energy released per unit area for each side of the
disk is (Shakura \& Sunyaev 1973)
\begin{equation}
F_g = {3\over 8\pi} {GM\dot M \over R^3} \hbox {J}(R)\;,
\end{equation}
where $\hbox {J}(R) = 1 - (R_i/R)^{1/2}$ and $R_i$ is the radius
of the last stable orbit, here taken to be $R_i = 3 R_g$, with $R_g \equiv 2GM/c^2$.
We assume that the temperature profile of the disk is given by
\begin{equation}
\theta \equiv kT= 0.1 \left({r\over r_o}\right)^n \hbox {J}^m\quad \hbox{\rm keV}\;.
\end{equation}
where $r_o$ is the radius of the disk at which $\theta = 0.1 \hbox {J}^m \approx
0.1$ keV. Using this profile and assuming {\it a priori} (but checked later
for self-consistency) that the radiative mechanism is saturated 
Comptonization (i.e., a Wien peak), we obtain the 
flux at earth (Misra et al. 1997):
\begin{equation}
f_E(E) = K E^2 \int^{r_o}_0 (\theta)^{-4} \exp(-E/\theta)\, 
r^{-2}\; dr\qquad \hbox{\rm photons}/\hbox{\rm sec}/\hbox{\rm cm}^2/\hbox{\rm keV} 
\; , \end{equation}
where $E$ is the photon energy in keV and
\begin{equation}
K = 5.59 \times 10^{27} \dot M D^{-2}\;,
\end{equation}
where $D$ is the distance to the source. This intrinsic spectrum is 
a function of $K$, $n$, $m$, and $r_o$. 

To this must be added a reflection component, which has been 
calculated using the appropriate Green's function (White, Lightman,
\& Zdziarski 1988), under the assumption that the reflecting medium 
is neutral and averaging over incident and observing angles.
The reflected component depends only on the intrinsic
spectrum and the solid angle subtended by the reflector, 
which is conveniently normalized by the factor $R_{ref} = \Omega/2\pi$.
Here $R_{ref} = 1$ would correspond to an angle averaged reflection spectrum 
due to a reflector which suspends a solid angle $2\pi$ with respect to the
X-ray source. Since the reflected spectrum depends on the 
incident and observing angles, the real solid angle could be different from
the best fit $R_{ref}$ which is treated here like a parameter. 
It should be noted that the shape of the reflected spectrum taking into account
the angle of observation is similar to the angle averaged
one, however there could be a difference in normalization (Magdziarz \& Zdziarski 1995).
Together with the earlier four parameters, the quantity $R_{ref}$ renders 
the final spectrum (which is used to fit the data) a function of 
five independent parameters.

Since the temperature profile of the transition region 
is obtained in this way, the unknown viscosity mechanism
does not need to be specified in order for us to obtain the disk 
structure. In principle, the disk equations can be inverted and 
the viscosity parameter ($\alpha$) can then be obtained as a 
function of radius. As shown by Misra et al. (1997), however,
this has only a limited value since the inferred $\alpha$ 
strongly depends on the details of the vertical radiative transfer. 
Here, we show that this is still the case, but since there 
are now four sets of data, it is possible to infer some
useful qualitative trends.  We will see that $\alpha$ is a 
function of radius and luminosity, which may be understood
in the context of its intrinsic dependence on the local parameters such
as density, height and temperature.

\section{Observations and Results }

Ginga observed Cygnus X-1 on several occasions, one of
which was 1991 June 6, during the OSSE viewing period 2. 
Gierlinski et al. (1997) have extracted four Ginga ($2-30$ keV)
data sets from this period and the corresponding 
near simultaneous OSSE observations to perform their
spectral fit. We are using the same set of 
observations for the present study. A systematic error 
of 1\% is included in each of the Ginga channels.

The OSSE data span the energy range from 50 to
1000 keV, and systematic errors are included.
These are computed from the
uncertainties in the low energy calibration and response
of the detectors, using in-orbit and prelaunch calibration
data. The energy-dependent systematic errors are expressed
as an uncertainty in the effective area in the OSSE 
response and are added in quadrature to the statistical
errors. They are most important at the lowest energies,
contributing an approximate 3\% uncertainty in effective 
area at 50 keV, decreasing to 0.3\% at 150 keV and above.
The details of these observations are given in Table 1
of Gierlinski et al. (1997).

We fit the four sets of data using the empirical model described
in the previous section. Following  Gierlinski et al. (1997), we
consider only data above 3 keV, so that the soft X-ray excess does
not effect the fitting. The relative normalization between Ginga
and OSSE is allowed to vary freely within $\pm$ 15 per cent. In all 
four cases, the best fit normalization falls within this range.
The spectrum is absorbed by a line of sight column density using the 
cross section given in Morrison \& McCammon (1983). A lower limit of
$5\times 10^{21}$ cm$^{-2}$, corresponding to galactic absorption, 
is imposed on all the optimizations. This is consistent with ASCA measurements by Ebisawa
et al. (1996). We model the iron line as 
a Gaussian, whose centroid energy is restricted to lie within 
the range $6.3-6.6$ keV and whose line width is $0.1-0.2$ keV, 
consistent with the ASCA measurements (Ebisawa et al. 1996).

Table 1 summarizes the results of our spectral fits. The $\chi^2$ 
in each case is comparable to or smaller than those obtained by
Gierlinski et al. (1997), who used a single temperature isotropic
external photon Comptonization model. We show in Figure 1 the unfolded
spectrum and the residuals for set 1. Unlike the single temperature
model, there are no strong systematic residuals and hence a second
component is not required here. Similar results were obtained for
the other three data sets. This solidly establishes the results indicated
earlier by Misra et al. (1997), who used non-simultaneous data. 
The spectral parameters (i.e., n, m and $r_o$) are similar for
all four data sets, which suggests that the spectral shape does not
vary strongly with luminosity, as was also pointed out by Gierlinski et al. (1997).

\begin{table}
\caption{Spectral parameters for the transition disk model}
\begin{tabular}{llllll}
\hline
Model &  Parameters & set-1 & set-2 & set-3 & set-4 \\
component & & &&& \\
\hline
abs & N$_{H}$ (10$^{22}$ cm$^{-2}$) & 0.5 & 0.57 $\pm$ 0.22 & 1.25 $\pm$ .22 & 1.1 $\pm$ .22 \\
Gaussian & E$_{Line}$ (keV) & 6.38 $\pm$ 0.1 & 6.44 $\pm$ 0.1 & 6.43 $\pm$ 0.11 & 6.31 $\pm$ 0.21\\
         & $\sigma$ (keV) & 0.2 & 0.2 & 0.2 & 0.2 \\
         & K$_{Line}$ & 9.06 $\pm$ 1.6  & 10.0 $\pm$ 1.7 & 5.59 $\pm$ 1.0 & 4.32 $\pm$ 1.3\\
	 & EW (eV) & 124 & 126 & 115 & 93.3 \\
transit  & $r_o$ & 114.9 $\pm$ 15.5 & 109.4 $\pm$ 10.3 & 155.9 $\pm$ 26.6 & 96.0 $\pm$ 17.6\\
         & $n$ & -2.50 $\pm$ 0.16  & -2.62 $\pm$ 0.24 & -2.18 $\pm$ 0.16 & -2.71$\pm$ 0.25 \\
         & $m$ & 0.87 $\pm$ 0.12 & 1.03 $\pm$ 0.18 & 0.62 $\pm$ 0.11 & 0.98 $\pm$ 0.2\\
         & $R_{ref}$ & 0.642 $\pm$ 0.06 & 0.644 $\pm$ 0.07 & 0.39 $\pm$ 0.06 & 0.513 $\pm$ 0.07\\
          & $K$ & 46.8 $\pm$ 0.61 & 54.0 $\pm$ 0.9 & 32.75 $\pm$ 0.4 & 29.98 $\pm$ 0.88\\
 & $K_{osse}$ & 0.92 $\pm$ .02 & 0.89 $\pm$ .02 & 0.99 $\pm$  0.02 & 0.97 $\pm$ .04\\
\hline
$\chi^2$ (dof) & & 58.96 (75) & 67.33 (75) & 60.59 (75) & 63.52 (75)\\
\hline
\end{tabular}

$^1$The normalization constants for the Gaussian line is in units of
10$^{-3}$ photons cm$^{-2}$ s$^{-1}$.

$K_{osse}$ : Normalization of the OSSE spectrum relative to the Ginga
spectrum.
\end{table}

To check the robustness of these results we have re-fitted the data
after fixing $n = -3.5$, which is different from the best fit value 
$n \approx -2.5$. The ensuing results are shown in Table 2.  Evidently,
the $\chi^2$ increased by about $\sim$ 10, but the reduced
$\chi^2$ is still $\sim 1$ and thus a temperature profile corresponding
to this set of parameters cannot be excluded. As we note in the next
section, these two sets of parameters give rise to different disk 
structures. However, since the $\Delta \chi^2$ is large, the latter
case should be considered an extreme case.
 
\begin{table}
\caption{Spectral parameters for the transition disk model with frozen $n = -3.5$}
\begin{tabular}{llllll}
\hline
Model &  Parameters & set-1 & set-2 & set-3 & set-4 \\
component & & &&& \\
\hline
abs & N$_{H}$ (10$^{22}$ cm$^{-2}$) & 1.18 $\pm$ 0.10 & 1.09 $\pm$ 0.1 & 2.27 $\pm$ 0.1 &  1.55 $\pm$ 0.10 \\
Gaussian & E$_{Line}$ (keV) & 6.49 $\pm$ 0.14 & 6.56 $\pm$ 0.15 & 6.6 & 6.39 $\pm$ 0.17\\
         & $\sigma$ (keV) & 0.2 & 0.2 & 0.2 & 0.2 \\
         & K$_{Line}$ & 6.5 $\pm$ 1.3 & 7.9 $\pm$ 1.5 & 3.0 $\pm$ 0.8 & 3.1 $\pm$ 0.8\\
	 & EW (eV) & 88 & 97.5 & 60 & 65  \\
transit  & $r_o$ & 63.43 $\pm$ 1.8 & 67.13 $\pm$ 2.1 & 63.35 $\pm$ 1.9 & 64.2 $\pm$ 3.7\\
         & $n$ & -3.5  & -3.5 & -3.5 & -3.5 \\
         & $m$ & 1.54 $\pm$ 0.07 & 1.69 $\pm$ 0.08 & 1.51 $\pm$ 0.07 & 1.59 $\pm$ 0.14\\
         & $R_{ref}$ & 0.849 $\pm$ 0.04 & 0.807 $\pm$ 0.04 & 0.702 $\pm$ 0.03 &  0.62 $\pm$ 0.06\\
          & $K$ & 50.2 $\pm$ 0.6 & 57.5 $\pm$ 0.7 & 35.38 $\pm$ 0.4 & 32.19 $\pm$ 0.68\\
 & $K_{osse}$ & 0.91 $\pm$ .02 & 0.87 $\pm$ .02 & 0.96 $\pm$  0.02 & 0.93 $\pm$ .04\\
\hline
$\chi^2$ (dof) & & 71.85 (76) & 72.99 (76) & 83.54  (76) & 67.65 (76)\\
\hline
\end{tabular}

$^1$The normalization constants for the Gaussian line is in units of
10$^{-3}$ photons cm$^{-2}$ s$^{-1}$.

$K_{osse}$ : Normalization of the OSSE spectrum relative to the Ginga
spectrum.
\end{table}

\section{The transition disk}

Using the best fit empirical temperature profile obtained in the previous
section, we can now derive the physical structure of the disk, subject
to a number of assumptions. The procedure was developed in detail 
by Misra et al. (1997), so we provide only a brief summary here.

The assumptions that will be checked for self-consistency
are 1) that the plasma cools via saturated bremsstrahlung self-Comptonization
and 2) that advection is not important. The structure equations are
those for hydrostatic equilibrium, the radiative transfer and radiative
cooling. These equations, together with Equation (1), completely determine 
the disk structure with the use of the best fit temperature profile. Note
that the torque equation generally used in standard disk solutions is
not needed here, since the necessary information is contained in
the temperature profile. Thus the unknown viscosity process does not need to be 
specified in this scheme. The handling of the radiative transfer is the most uncertain
among these equations and is accounted for with the expression 
\begin{equation}
F_g = B_1 {c P_r\over \tau}\;,
\end{equation}
where $F_g$ is the  energy flux, $P_r$ is the radiation pressure
and $\tau$ is the scattering optical depth. The uncertainty 
is represented by the factor $B_1$, which is taken to be of order unity.
In the radiative cooling equation, the bremsstrahlung rate is multiplied
by the Compton Amplification factor (Svensson 1983).
Using the proton electron Coulomb energy transfer we take into account
the possibility that the proton temperature may be different from that
of the electrons. The total pressure is taken to be the sum of
the radiation and gas pressures.  

The normalization parameter $K$ (Eq. 4) may be directly related to the
mass accretion rate (Misra et al. 1997):
\begin{equation}
\dot M = 5.3 \times 10^{17} \left({K\over 50}\right) 
\left({D\over 2.5 \hbox{\rm kpc}}\right)^2\quad \hbox{\rm grams}/\hbox{\rm sec}\;.  
\end{equation}
We take the distance to the source to be $2.5$ kpc and assume that
the black hole mass is ten solar masses.

In Figure 2, we show the variation of temperature, scattering optical
depth and pressure with radius for set 1, in which $B_1 = 0.25$. 
These disk characteristics are qualitatively similar to those
inferred earlier by Misra et al. (1997). In Figure 3, the 
effective optical depth $\tau_{eff}$ is plotted as a function 
of radius (appearing in the lower right hand corner). In the
same plot, the Compton $y$ parameter $y\equiv ({4kT_e/ m_ec^2}) \tau^2 $ is
shown as a function of radius. The fact that $\tau_{eff} < 1$ and $y \sim
1$ justifies the assumption that the radiative mechanism is Comptonized
bremsstrahlung.  To check the self-consistency of our second assumption
(regarding the insignificance of advection), we have also calculated
the ratio of the pressure to the number density times
the virial temperature ($\approx m_pc^2/r$):
\begin{equation}
{P_a/P} = {\rho c^2\over r}{1\over P}\;,
\end{equation}
and have plotted it in Figure 3.  Since this ratio is always much 
greater than one, advection is not important and the disk is 
geometrically thin. Analogous self-consistency checks have been 
carried out for the other three observational sets as well.

Following Misra et al. (1997), we have next calculated the 
viscosity parameter $\alpha$ as a function of radius
from the inverted disk equations. Figure 4 shows this
quantity corresponding to the first data set for two values
of $B_1$ and $n$.  Clearly, $\alpha$ depends rather strongly
on the actual value of $B_1$, and hence the radiative transfer, as
may be inferred from a comparison of the solid and dotted curves
in this figure.  It also depends in an important way on $n$, as
may be seen by inspecting the dashed curve in Figure 4.
Since the data do not discriminate strongly among the variations
of $\alpha$ with radius, it is difficult to make concrete statements 
about the viscosity based on these results.

Nonetheless, with four sets of data corresponding to different
luminosities, it is still possible to discern qualitative trends. Figure
5 shows the variation of $\alpha$ with radius for the best fit 
temperature profile in each of the four data sets (solid curves). 
The viscosity parameter $\alpha$ seems to vary inversely with 
the luminosity, which itself scales with the accretion rate. 
However, $\alpha$ should in principle depend only on 
local parameters such as height, density and temperature. We have
found that the following empirical ``fit'' for $\alpha$ in terms of these 
parameters, reproduces the variation quite well:
\begin{equation}
\alpha (\Xi, T) = 2.65\, \left({T\over 10^8 \hbox{\rm K}}\right)^{-0.85}\; 
\left({\Xi\over 10\;\hbox{\rm g}\;{\hbox{\rm cm}}^{-2}}\right)^{-2.8}\;,
\end{equation} 
where $\Xi$ is the column density.
This empirical fit is shown (dotted curve) in Figure 5. Care should be
taken when interpreting this result since $\alpha$ depends sensitively
on the uncertainty in the radiative transfer. Moreover,
the fit itself may not be unique. We are showing this here primarily
to demonstrate that it is possible to obtain a functional form for 
$\alpha$ that depends only on the local parameters.

\section{Summary}

We have produced a fit to four sets of simultaneous broad band 
data for Cygnus X-1 within the context of the transition disk model.
These data were obtained with observations by OSSE ($50-500$ keV)
and Ginga ($3-30$ keV).  We have established rather strongly
that the transition disk model fits the data better
than a pure soft-photon Comptonization model, as suggested earlier by
Misra et al. (1997) based on non-simultaneous data. In particular the
$\chi^2$ value for the transition disk model is in each case
smaller than or comparable to that of the isotropic Comptonization 
model fit to the same data (Gierlinski et al. 1997). 
In addition, unlike the soft-photon Comptonization
model, there are no systematic residuals for the transition disk fit. 
The transition disk is described by four parameters as compared to
the isotropic Comptonization model which has three. However, for
the model presented here no additional ad hoc component, with
two extra parameters (Gierlinski et al. 1997) are needed.

We have also determined the disk structure from the best fit 
temperature profiles, and through self-consistency checks, 
we showed that our initial assumptions were valid. 
The viscosity parameter $\alpha$ inferred from the calculated
disk structure was found to depend on the details of the vertical
radiative transfer and the spectral fitting. Thus, 
it is not yet possible to make concrete statements 
regarding it's value or specific tendencies. However, 
since we were able to analyze the data at four different
epochs (each corresponding to a different luminosity),
we derived the qualitative variation of $\alpha$ with 
radius and luminosity, which we accounted for in terms
of its dependence on the local disk structure parameters,
such as density, temperature and height.

\acknowledgments

The authors thank A. A. Zdziarski and M. Gierlinski
for providing us with the reduced data from OSSE and Ginga. RM
and VRC thank A. Kembhavi for useful discussions. 
VRC acknowledges IUCAA for its hospitality. 
This work was partially supported by NASA grant NAG 5-3075.

\clearpage
 
\figcaption{The deconvolved spectrum of Cygnus X-1 corresponding
to the first data set, obtained from simultaneous observations
with Ginga and OSSE. The contributions from individual model
components, viz., the iron line and reflection, are shown 
separately. The residuals to the model fit are shown in the
lower panel as a contribution to the $\chi^2$.\label{fig1}} 

\figcaption{(a) The temperature profile in the disk for the
best fit values in data set 1, and $B_1 = 0.25$. Solid curve:
electron temperature; dotted curve: proton temperature.
(b) The scattering optical depth ($\tau = n_p\sigma_T H$) as a function 
of radius corresponding to the temperature profile in (a) and $B_1 = 0.25$. 
(c) The mass density as a function of radius corresponding to the 
temperature profile in (a) and $B_1 = 0.25$.
(d) Pressure as a function of radius corresponding to the temperature 
profile in (a) and $B_1 = 0.25$. Solid curve: Radiation pressure; Dotted
curve: Gas pressure.\label{fig2}}

\figcaption{The effective optical depth ($\tau_{eff}$), the Compton $y$ 
parameter and the ratio $P_a/P$ as functions of radius for data set 1. \label{fig3}}

\figcaption{The viscosity parameter $\alpha$ as a function of radius for 
data set 1. Solid curve: $B_1 = 0.25$; Dashed curve: $B_1 = 0.5$. The
dotted curve is for a temperature profile calculated with $n=-3.5$. \label{fig4}}

\figcaption{The viscosity parameter $\alpha$ (solid curve) as a function 
of radius for data sets 1, 2, 3, and 4. Dotted curve: empirical fit given 
by Equation (8) in the text.
\label{fig5}}

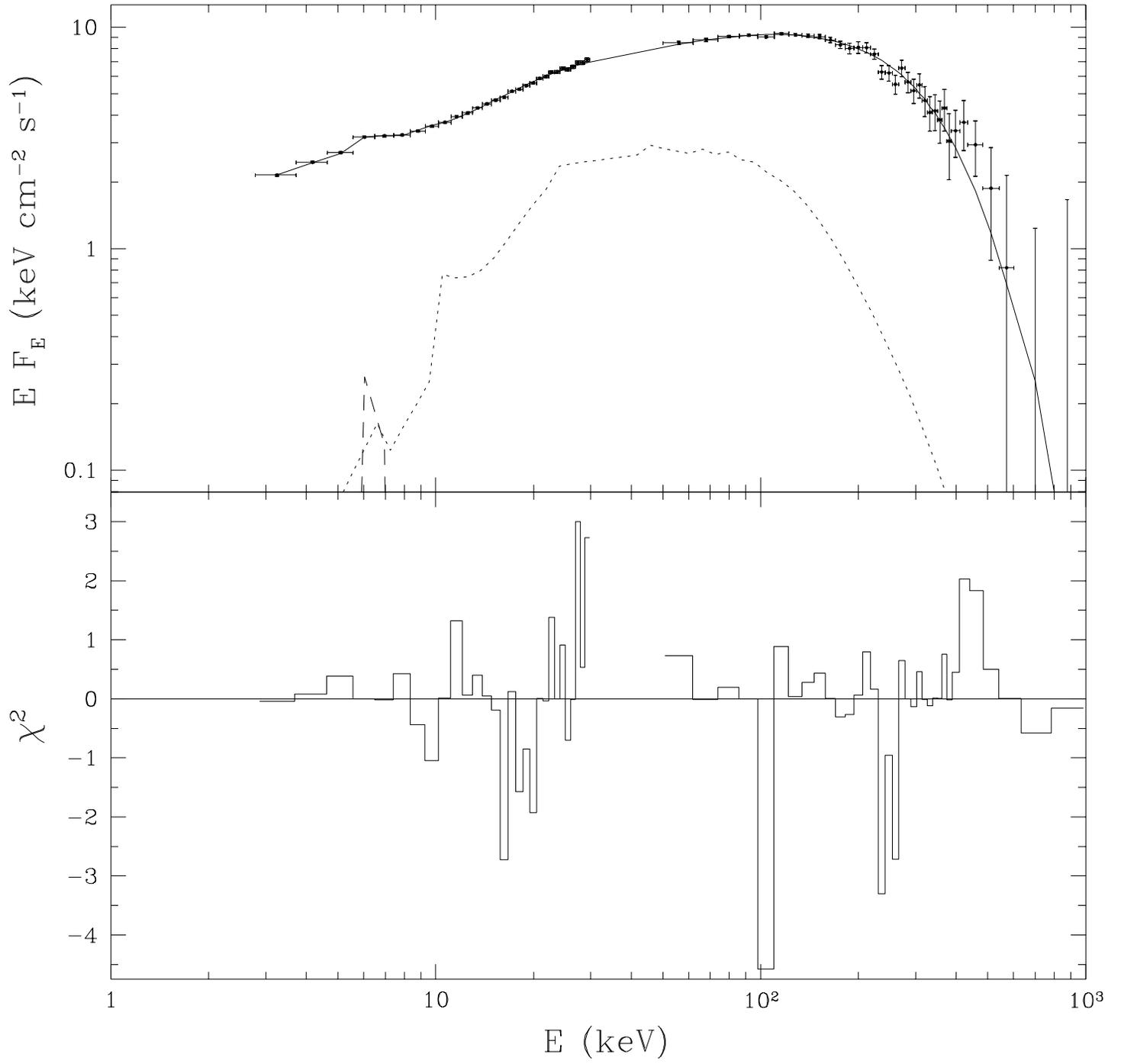
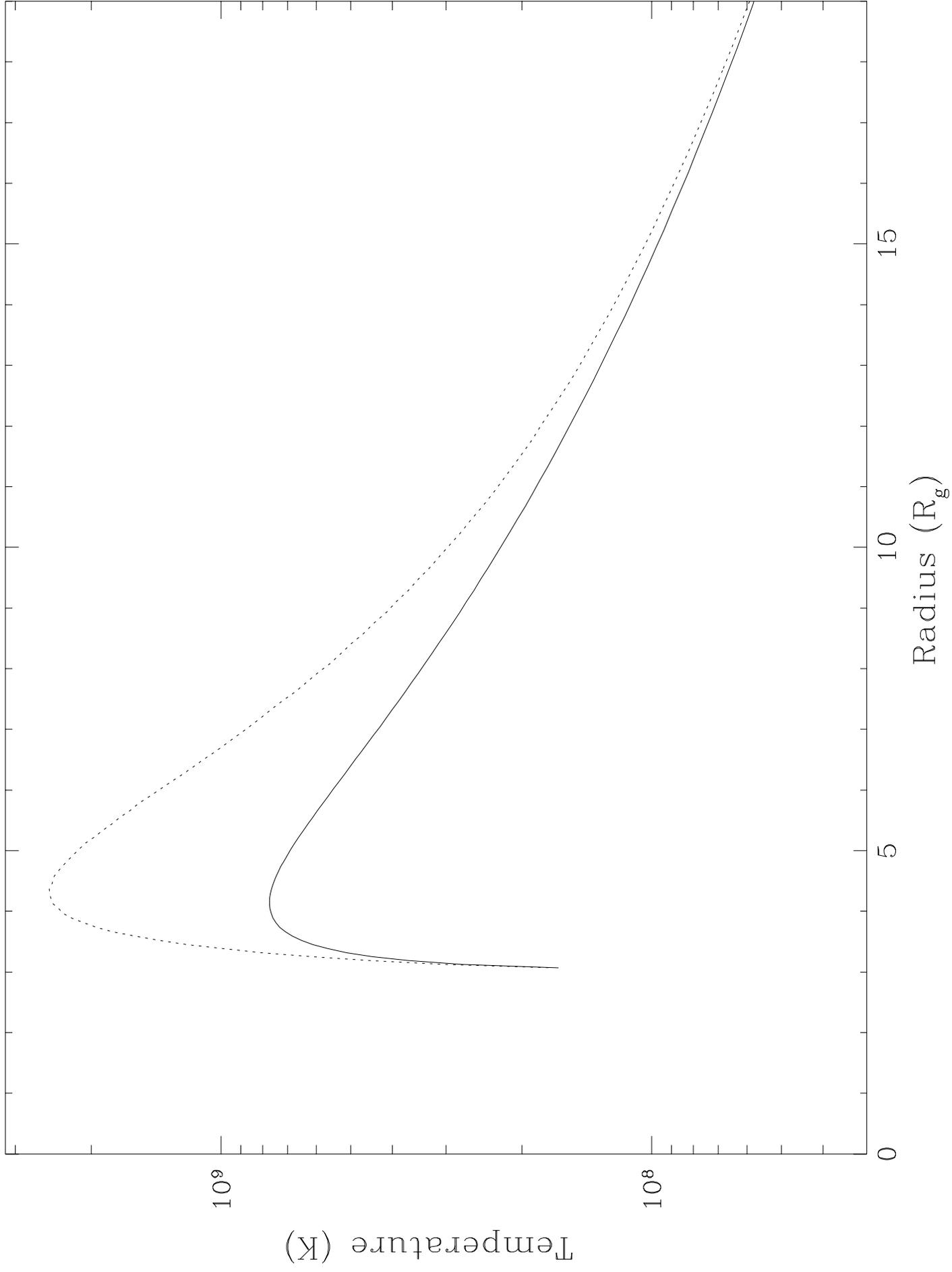
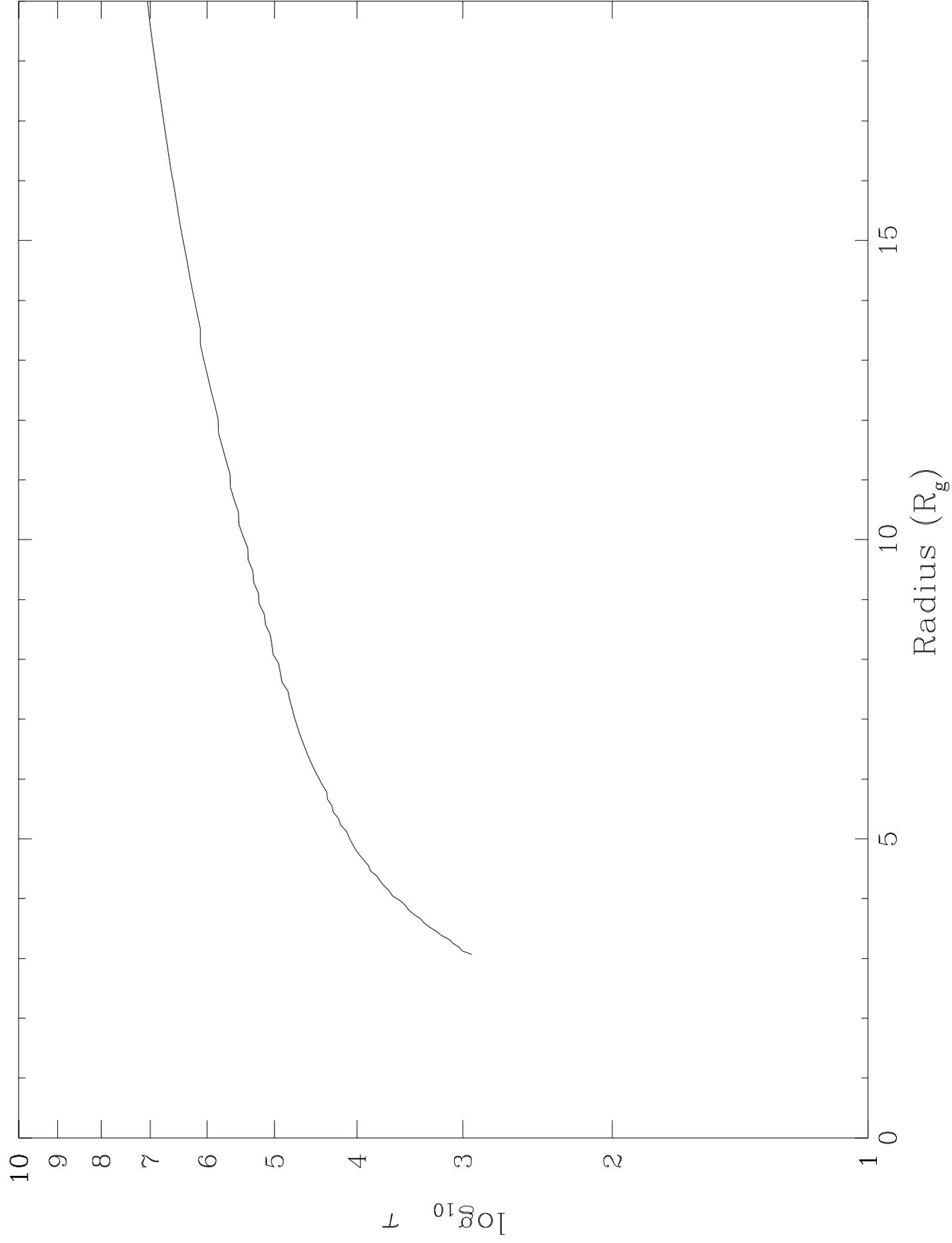
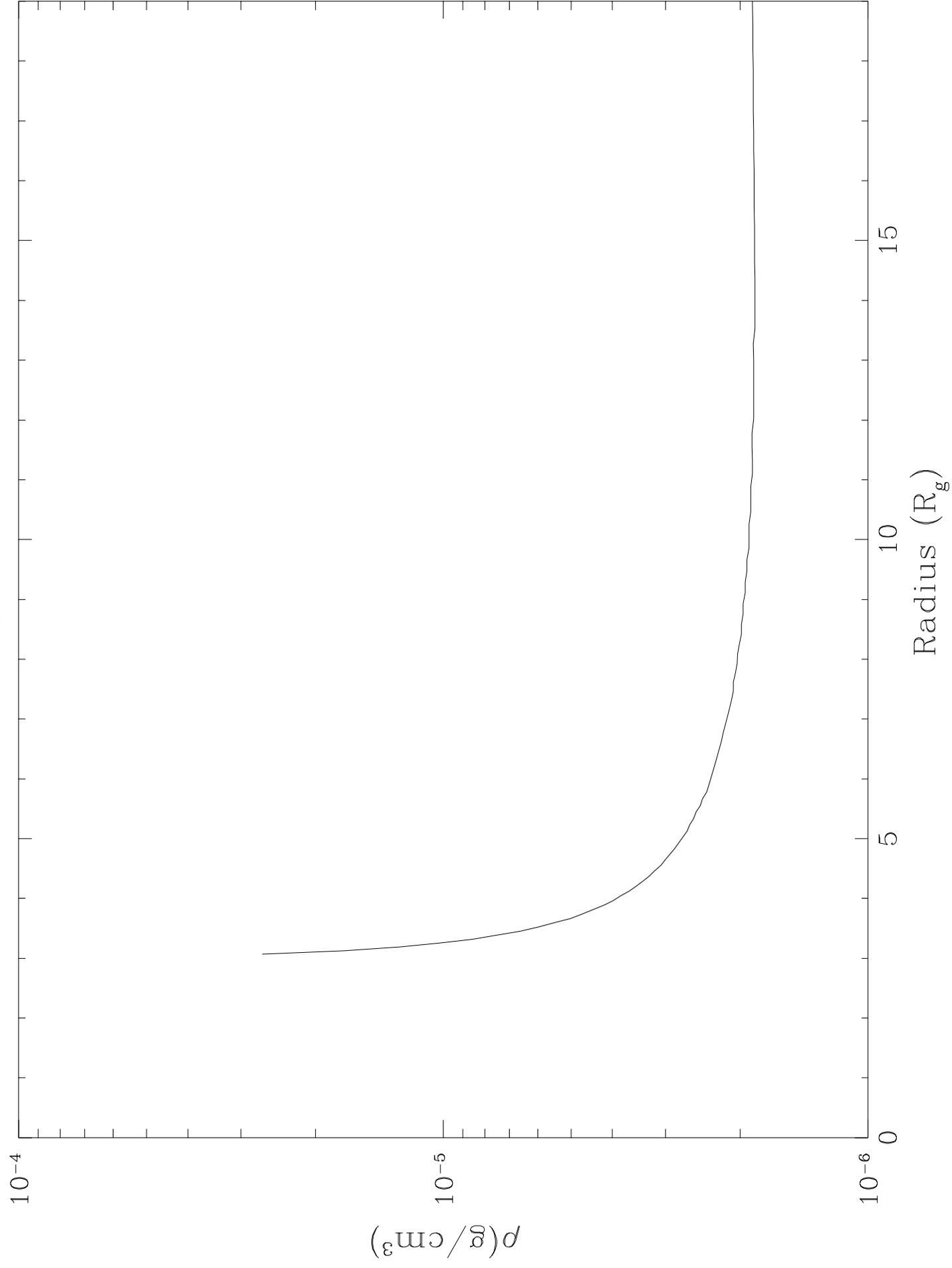
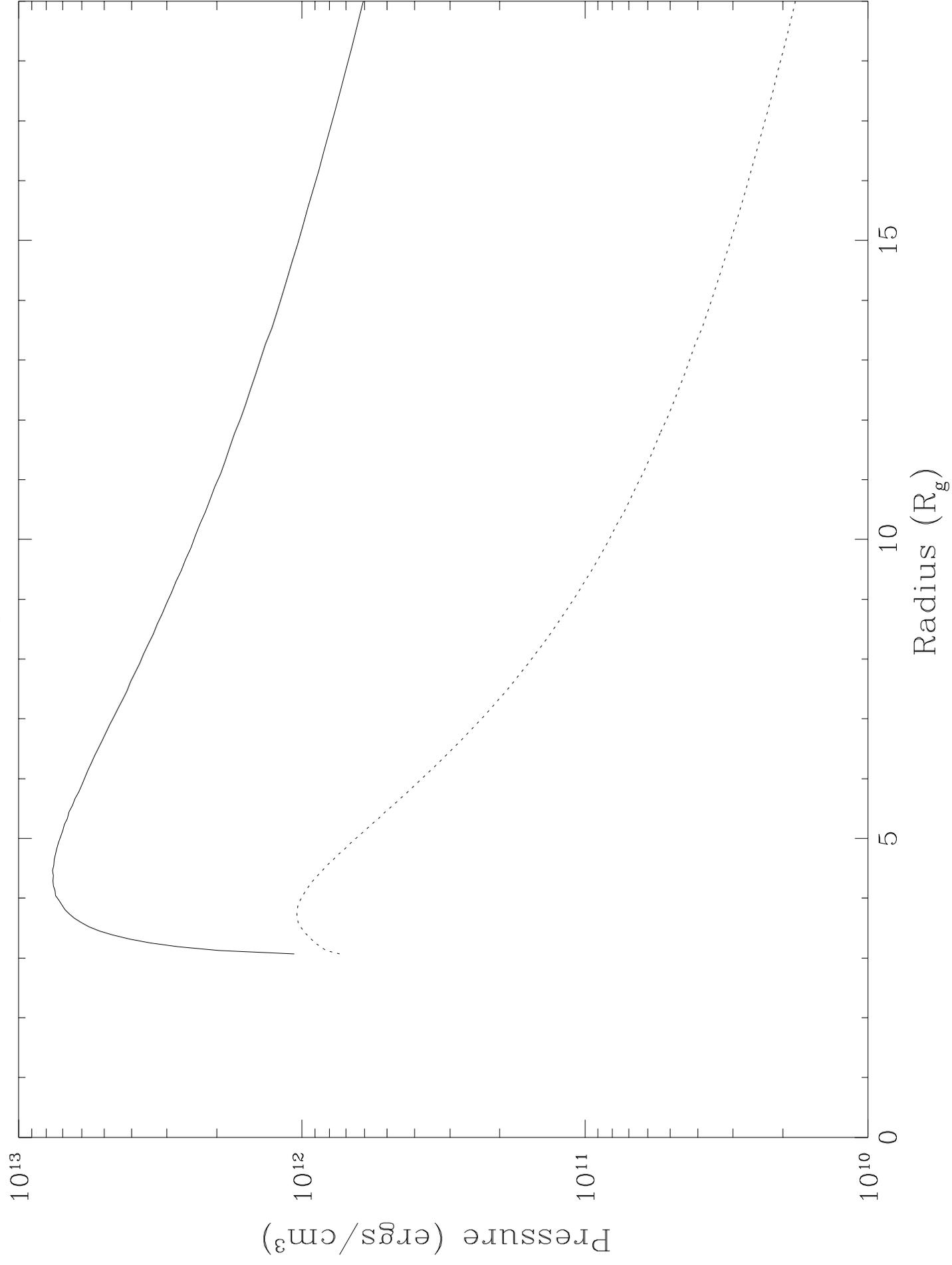
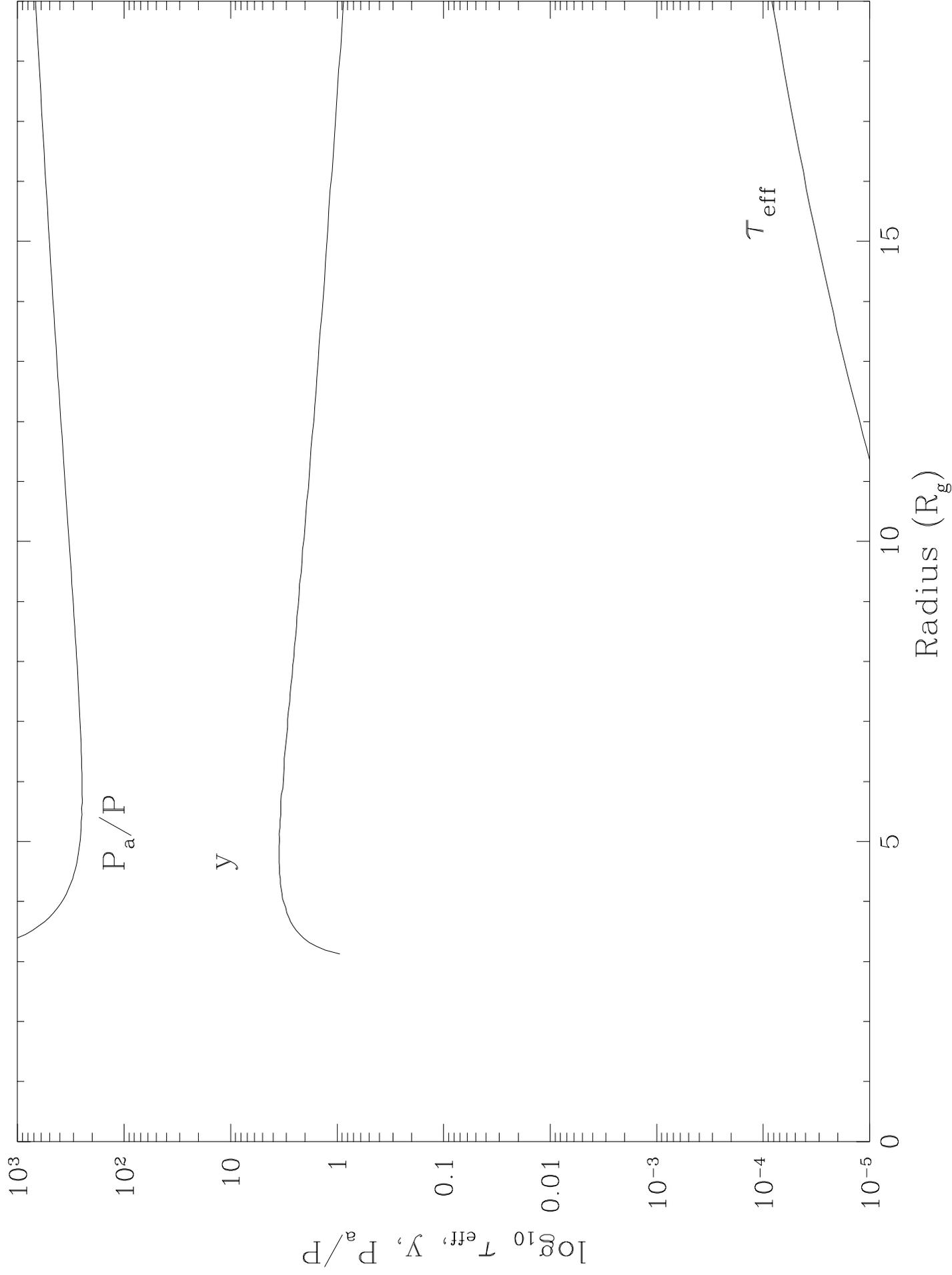
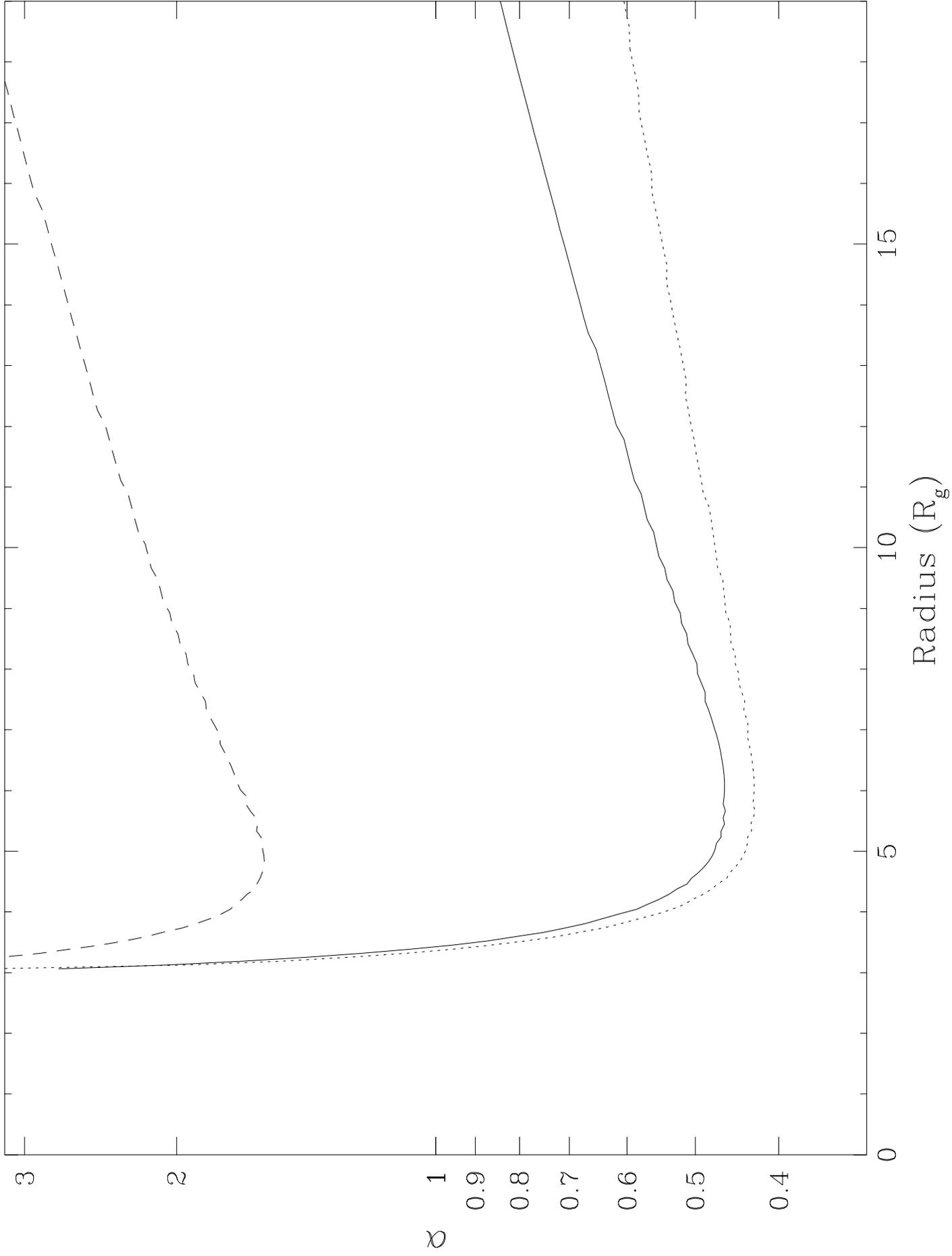
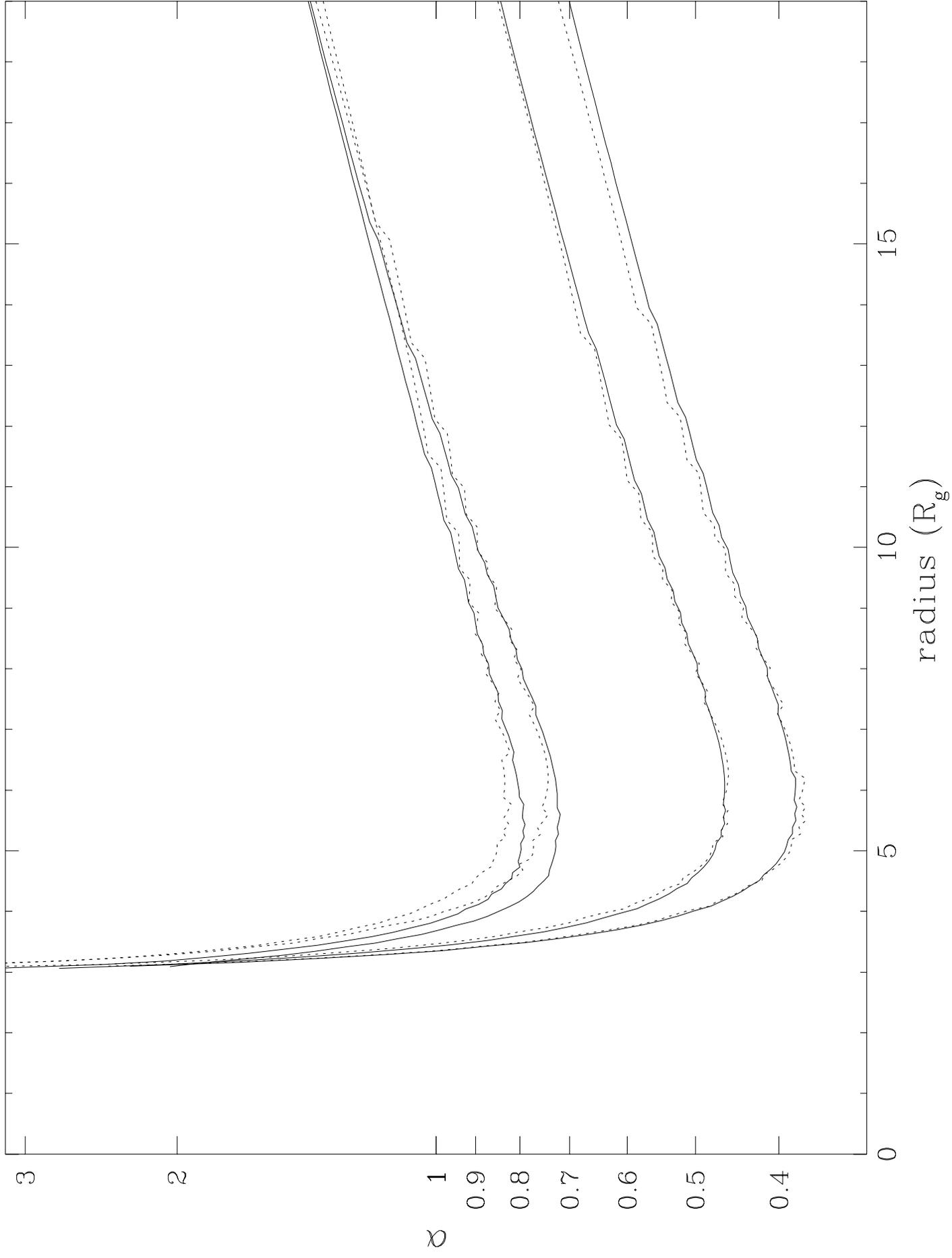
\end{document}